\documentclass[preprint,3p]{elsarticle}
\usepackage{natbib}
\usepackage{amssymb}
\usepackage{amsmath}
\usepackage{graphicx}
\usepackage{subfigure}

\begin{document}
\begin{frontmatter}

\title{A damage model based on failure threshold weakening}

\author[phys]{Joseph~D.~Gran\corref{cor1}}
\ead{gran@student.physics.ucdavis.edu}

\author[phys,geo,stf]{John~B.~Rundle}
\ead{rundle@physics.ucdavis.edu}

\author[geo]{Donald~L.~Turcotte}
\ead{dlturcotte@ucdavis.edu}

\author[phys]{James~R.~Holliday}
\ead{holliday@physics.ucdavis.edu}

\author[bu]{William~Klein}
\ead{klein@bu.edu}

\cortext[cor1]{Corresponding author}
\address[phys]{Department of Physics, University of California, Davis, Davis CA, 95616}
\address[geo]{Department of Geology, University of California, Davis, Davis CA, 95616}
\address[stf]{Santa Fe Institute, Santa Fe, NM 87501}
\address[bu]{Department of Physics, Boston University, Boston, MA 02215}

\date{\today}

\begin{abstract}
A variety of studies have modeled the physics of material deformation and damage as examples of generalized phase transitions, involving either critical phenomena or spinodal nucleation.  Here we study a model for frictional sliding with long range interactions and recurrent damage that is parameterized by a process of damage and partial healing during sliding.  We introduce a failure threshold weakening parameter into the cellular-automaton slider-block model which allows blocks to fail at a reduced failure threshold for all subsequent failures during an event. We show that a critical point is reached beyond which the probability of a system-wide event scales with this weakening parameter.  We provide a mapping to the percolation transition, and show that the values of the scaling exponents approach the values for mean-field percolation (spinodal nucleation) as lattice size $L$ is increased for fixed $R$.  We also examine the effect of the weakening parameter on the frequency-magnitude scaling relationship and the ergodic behavior of the model.
\end{abstract}

\begin{keyword}
Damage; Slider-block model; Scaling; Percolation; Critical point
\end{keyword}

\end{frontmatter}

\section{Introduction}

The physics of progressive material damage and healing under load has been studied using models that incorporate both failure thresholds and load sharing.  Models of this class include slider-block models and fiber-bundle models.   In both models, external loads are applied to a ``loader plate'' to which the blocks or fibers are attached.  

One type of avalanche model, using sliding blocks have been used to study the physics of frictional sliding, earthquakes~\citep{Burr1967,Rund1977,Brow1991}, neural networks~\citep{Hopf1995}, as well as other driven threshold systems in nature.  A variety of phenomena have been studied using these models, including aspects of scaling and critical phenomena~\citep{Carl1989Prop,Rund1989,Rund1996,Klei1997,Rund1997}, as well as nucleation near the spinodal.  A primary focus has been placed upon understanding the origins of the frequency-size relation seen in earthquake systems~\citep{Burr1967,Rund1977,Brow1991,Hopf1995,Carl1989Prop,Rund1989,Rund1996,Klei1997,Rund1997}.  Slider-block models are closely related to other types of models for distributed failure, including fiber-bundle models~\citep{Nech2005,Prid2002,Turc2003,Virg2007} and fuse models~\citep{Zapp1997} for the study of material damage and degradation and forest fire models~\citep{Dros1992}.

Models for damage usually involve a system with a large number N of brittle elements such as fiber-bundles~\citep{Prid2002,Newm2001,Phoe2009}. Often the elements have a statistical distribution of strengths.  An external load is applied  to the system, and several failure modes are typically observed.  In some models, the fibers are completely brittle, so that failure occurs when the load on the fiber reaches the fiber strength.  In other models, the strength of the element can degrade and weaken with time when the element load reaches a weakening value, with the time to failure dependent on the amount by which the stress exceeds the weakening value~\citep{Curt1997b,Prid2002,Turc2003,Virg2007}.  In these models, an upper brittle strength value allows immediate failure when the stress on the element equals or exceeds the brittle strength.  In these models, a damage variable $\alpha$ can be defined as the fraction of elements or fibers that have failed.

Load sharing is an important property of both damage models and slider-block models.  In the classical hierarchical fiber-bundle models~\citep{Dani1945,Pier1926,Newm1995}, all fibers are connected to end plates to which the external tensional stress   is applied.  When a fiber fails, its load is transferred in equal parts to all the other intact fibers through the end plates.  Other details of fiber-bundle models have been discussed extensively in the literature~\citep{Lei2007, Lyak2005, Lyak2001, Main2000, Tous2002-1, Tous2002-2, Tous2002-3}.

In the slider-block model, a loader plate is connected by springs to an array of blocks sliding on a frictional surface, with the blocks interconnected by coupling springs~\citep{Burr1967,Rund1977}.  The loader plate  is typically driven either by a constant applied force $F$ or by a constant applied velocity V.  

In the earliest slider-block models, a failure threshold (static friction threshold) was prescribed along with a residual or arrest stress level.  Stress transfer in a slider-block model can occur through the coupling springs as well as through the loader plate. 

Healing is another physical process that occurs in some models of recurrent damage~\citep{Diet1979}, and can be observed in the laboratory and in nature.  When an element fails, it transitions suddenly to a new state (``ruptured state'').  For a process such as cracking under tensile load, separation of the crack surfaces generally precludes healing.  However, for shear failure, the crack surfaces typically remain in close proximity, and healing re-strengthening is possible under certain load conditions~\citep{Diet1979}.  This is the case, for example, in sliding friction experiments, where strength increases as $\log(\Delta t)$ following a slip event, where $\Delta t$ is the time interval since the slip. Presumably, the same process operates on natural faults and shearing surfaces, since sudden stick slip events are observed to reoccur regularly on these surfaces as earthquakes~\citep{Hami2006,Katz2004}, and therefore healing must occur.

\textit{To summarize our results:} We study a modified slider-block model for damage and failure in materials leading to a family of scaling exponents that can be measured in simulations.  Our model differs from a traditional slider-block model in that we have included failure threshold weakening.  Subsequent failures of a block during an event, will occur at a reduced failure threshold, parameterized by $w$.  The failure threshold weakening parameter $w$ represents the percentage reduction of the failure threshold after a block's initial slip.  We investigate the behavior of the model as $w$ is varied and show that above a critical value, $w_c$, system-wide events occur regularly.  The probability of a system-wide event scales with the distance from the critical point, $w-w_c$.  We examine the finite size effects and show in the limit of $L \rightarrow\infty$ the scaling exponent $\beta$, which characterizes the probability of system-wide events, approaches the corresponding order parameter exponent of mean-field percolation.

This paper is organized as follows.  In section 2 we discuss details of the model we use for simulations.  In section 3 we present the results of our simulations, and in section 4 we discuss the mapping to percolation theory.  In particular, we compare values of scaling exponents we obtain from the simulations to values for mean-field percolation and spinodal nucleation.  We also place our results into the context of other recent work on similar models, with particular attention to the question of punctuated ergodicity~\citep{Thir1990,Klei1997,Tiam2003,Seri2009} which has been observed in both models as well as in natural earthquake fault systems.

\section{Model Definition}
The model we investigate here is a cellular automata slider-block model. We incorporate an idea for the physics of weakening and healing proposed in a different context~\citep{BenZ1999}, which was a model for damage evolution in a continuum model of the earth's crust.  Inasmuch as cellular automata slider-block models are simple systems in which to examine the physics of correlation and scaling, we have constructed a modified form of the usual slider-block model which has loader or pulling springs, connecting or interacting springs, and fixed failure and residual strengths.  

Our model includes loader and coupling springs as well as additional ``weakening'' and ``healing''  properties~\citep{BenZ1999}.  By suitably tuning the threshold weakening, we find intervals in which a quasi-periodic earthquake cycle is observed that are dominated by large ``characteristic'' or ``nucleation'' events interspersed with time intervals in which a Gutenberg-Richter, or scaling distribution of events is observed.  In the following, we investigate and extend this model further, focusing in particular on the values of scaling exponents that are observed.  It will be seen that the amount of threshold weakening can be described by means of a parameter $w$  which appears to act as a relevant scaling field in the sense of nucleation and critical phenomena.  

Many dynamical models, for example forest fires~\citep{Dros1992}, display clusters of sites that are typically mapped onto percolation clusters, therefore our focus in this paper is on determining how the scaling exponents for the slider-block clusters are related to the scaling exponents typical of percolation theory.  In particular, as we are interested in slider-block models with long range interactions, approaching mean-field, we examine mean-field percolation models in spatial dimensions $d \geq 6$~\citep{Stau1994}, the upper critical dimension.

An example of mean-field percolation is provided by the Bethe lattice (Cayley Tree), which is equivalent to percolation in $d \rightarrow \infty$~\citep{Stau1994}).  In that problem, two of the critical exponents have the values~(\citet[see][table 2]{Stau1994}):   Order parameter exponent $\beta = 1$, and Fisher exponent $\tau = 2.5$.  The remaining critical exponents can be recovered from the well-known scaling relations~\citep{Stau1994}.  We note that for site percolation, the order parameter $P(p)$ can be defined as the probability that a spanning or percolating cluster is present, i.e., a cluster whose bonds connect all sides of the lattice.

\textit{Scaling in the slider-block model:} In the model we consider here, each of the individual avalanches or clusters of ``microscopic'' blocks, which can be quite large, is considered to be a cooperative ``macroscopic'' slip event, or ``earthquake''.  Many studies have shown that if the frequency of events is plotted as a function of the number of event size (e.g. blocks in a cluster), the result is a power law of the form~\citep{Carl1989Prop,Rund1996,Dros1992,Klei2000,Gaba2003}:
\begin{equation}
N = \frac{N_0}{A^{\tau-1}}e^{-\varepsilon A^{\sigma}}
\label{eqn1}
\end{equation}
where $\tau$ and $\sigma$ are critical exponents. Here $N$ is the number of clusters in the simulation with area $A$ and $\varepsilon$ is a scaling field.  In percolation theory, equation \eqref{eqn1} is called the Fisher-Stauffer droplet model~\citep{Stau1994}, and $\varepsilon \propto (p - p_{c})$, where $p$ is the occupation probability and $p_{c}$ is the value at the percolation threshold (critical point).  In earthquake seismology, a similar scaling equation is called the Gutenberg-Richter frequency-magnitude relation~\citep{Burr1967,Rund1977,Hopf1995,Carl1989Prop,Klei1997}.   The dynamics of the model are therefore reflected in the clusters of failed blocks and their statistics. 

\begin{figure}
\centering
\mbox{
\subfigure{\includegraphics[width = 0.45\columnwidth]{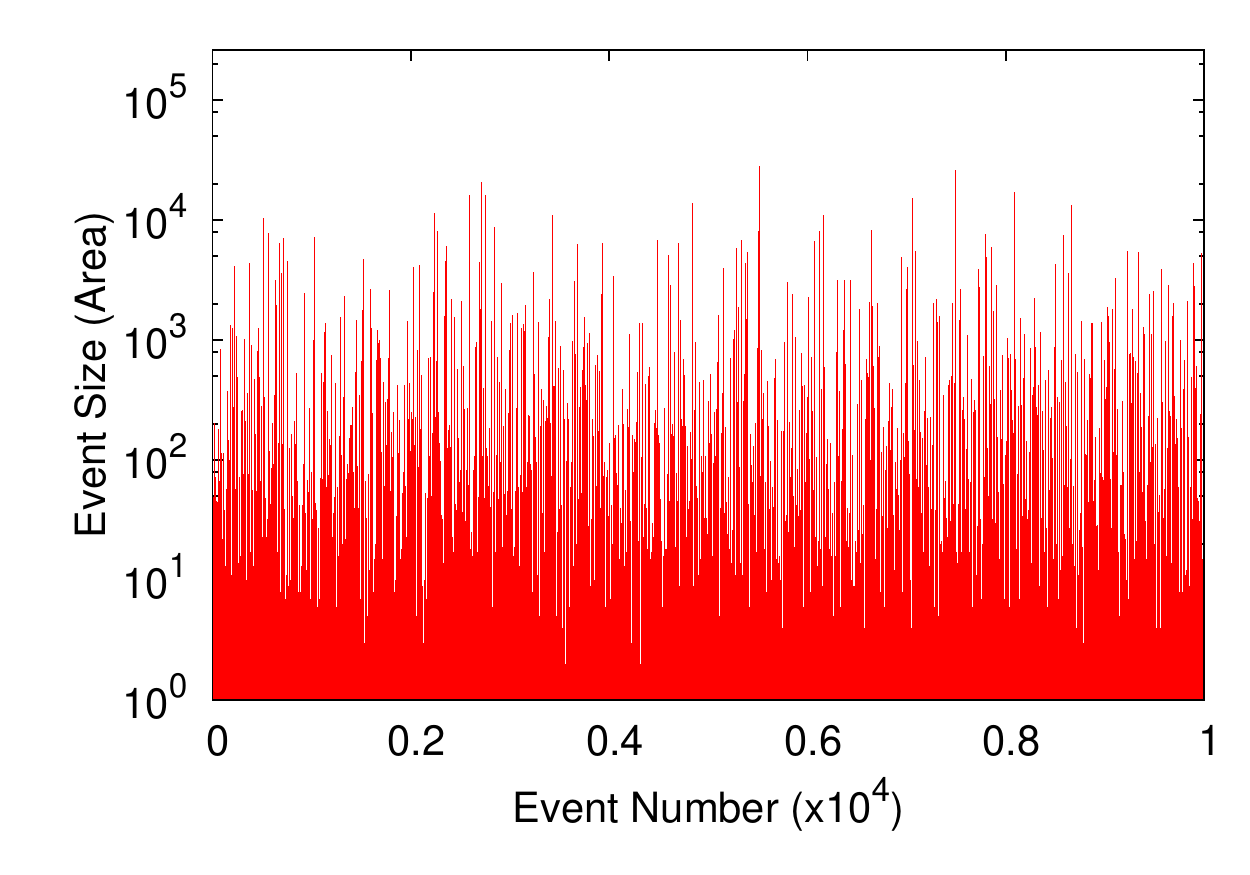}}
\quad
\subfigure{\includegraphics[width = 0.45\columnwidth]{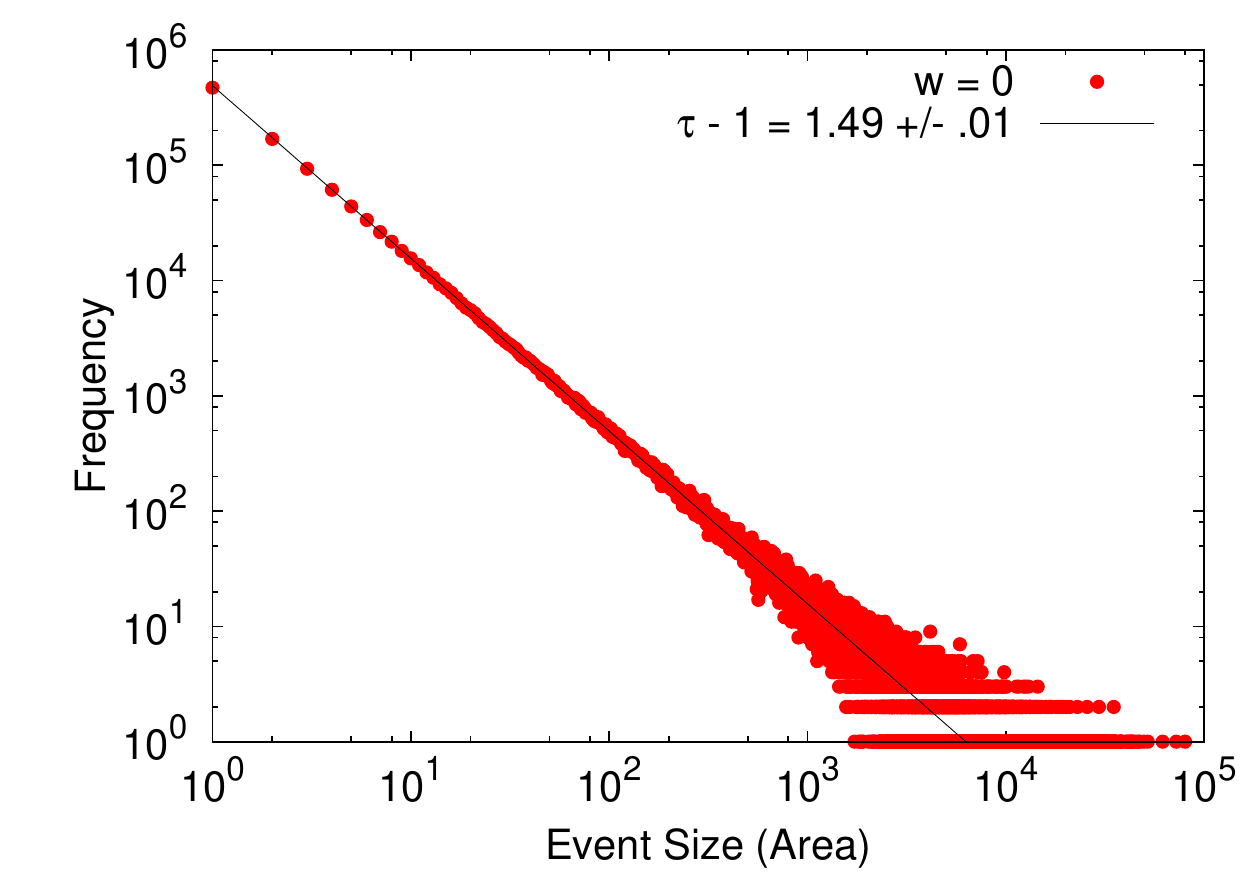}}}
\caption{
(a)~A sequence of event areas $A$ are given as a function of event number (natural time) for zero weakening ($w=0$). A small segment of the total sequence is shown. The largest events are semi-periodic but are much smaller than the lattice area $L^{2} = 512^{2} = 262,144$. The other parameters of the model run are: $\sigma_F =100$, $\sigma_R= 25 \pm 25$, $R = 15$ ($q = 960$), $K_C/K_L = 0.10$.
(b)~Frequency-magnitude statistics for weakening parameter $w=0$. The frequency of events of area $A$, is plotted as a function of area $A$.  The smaller events correlate well with the mean-field scaling relation taking $\tau - 1 = 1.49 \pm .05$.
}
\label{Events}
\end{figure}

\subsection{Base Model}
We first introduce the standard cellular automata slider-block model.  Our model consists of a 2d lattice of blocks sliding on a frictional surface. Our simulations are carried out on $LxL$ square lattices ranging in size from $L=256$ to $L=1024$, using periodic boundary conditions.  Each of the blocks is coupled to a loader plate by means of a loader spring having spring constant $K_L$.  Each block is connected to $q$ neighbor blocks by means of coupling springs having uniform spring constants $K_C$. The ``total spring constant'' is $qK_C + K_L$. The neighbor blocks form a square region with sides of length $2R+1$. Thus $q = (2R+1)^2 -1$, where $R$ is the range of interaction. Under rather general assumptions, it can be shown that the system can be regarded as approaching long range interactions $R\rightarrow \infty$, the mean-field limit, as $qK_C\rightarrow \infty$~\citep{Klei2000}. Typical slider-block model dynamics occur when the loader plate is advanced in displacement until the force or stress $\sigma$ on a single block meets or exceeds a failure threshold value $\sigma_F$, at which point the block is advanced a slip distance corresponding to a decrease in stress to the residual value.  The residual stress value is drawn from a uniform distribution centered at $\sigma_R$ with a spread of $\delta\sigma_R$ for each slip event.  This small random component added to the slip distance has the effect of thermalizing the system~\citep{Rund1995}.

Once the first block slips, stress is transferred to its $q$ neighbors by means of the coupling springs, and in turn, other blocks may slip as well.  The result is a cascade of slipping or failing blocks, often called an avalanche or event.  We note that we use ``zero-velocity'' dynamics, meaning that the loader plate is advanced just enough so that only a single block is at the failure threshold at the initiation of an event and the loader plate is held fixed in location until the event has ceased.  The loader plate is then advanced until the next event begins and the process is repeated.  These events share much in common with avalanche phenomena in other physical systems, including sandpiles, neural networks, and electronic devices such as diodes.  

We illustrate the behavior of our base model in Fig.~\ref{Events}.  In Fig.~\ref{Events}(a) we plot the size of sequential avalanches $A$ as a function of event number (natural time). The size of an avalanche is characterized by the total area (number of blocks) that slips during the event.  We give results for $10,000$ events from the middle of a simulation that included $10^6$ events. In this simulation the system size is $L = 512$, the interaction range is $R = 15$ ($q = 960$), $K_C/K_L = 0.1$, $\sigma_F = 100$ and $\sigma_R = 25 \pm 25$.  We see a wide range of rupture areas, the largest events occurring at relatively regular intervals. No system-wide events occur and the largest events, $A \approx 25000$ are much smaller than the system size $L^{2} = 262,144$. In Fig.~\ref{Events}(b) we plot the probability density function for the distribution of event sizes over the whole simulation run, the Gutenburg-Richter distribution. The frequency of events of area $A$ is plotted as a function of area $A$. By suitably tuning the system parameters $K_C/K_L$ and $R$ the model shows a mean-field scaling relationship with exponent $\tau-1 = 1.5$~\citep{Klei2000}.

\begin{figure}
\centering
\mbox{
\subfigure{\includegraphics[width = 0.45\columnwidth]{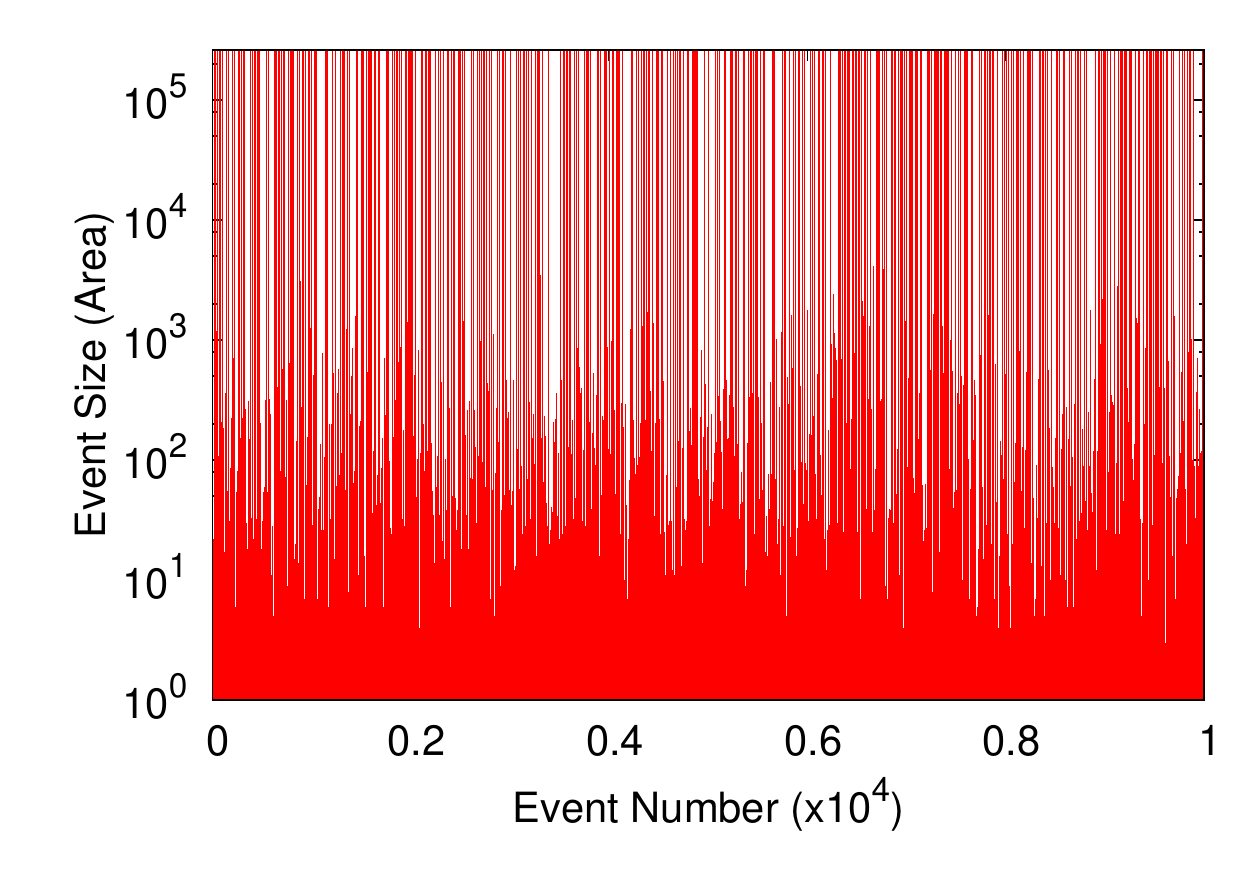}}
\quad
\subfigure{\includegraphics[width = 0.45\columnwidth]{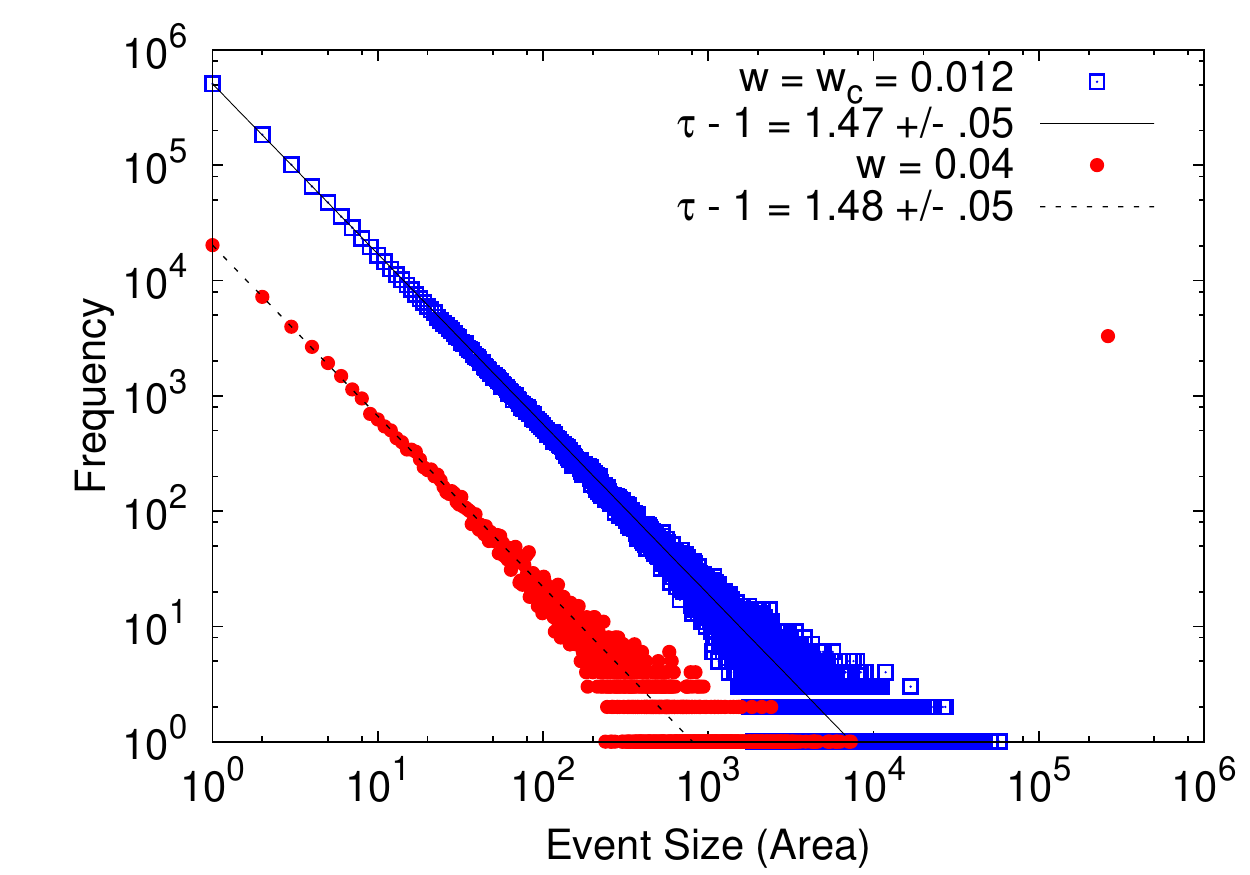}}}
\caption{
(a)~A sequence of event areas $A$ are given as a function of event number $N$ (natural time) for weakening above the critical value ($w>w_c$). A small segment of the total sequence is shown. System-wide events $A=L^{2} = 512^{2} = 262,144$ occur regularly.  The other parameters of the model run are: $\sigma_F =100$, $\sigma_R= 25 \pm 25$, $R = 15$ ($q = 960$), $K_C/K_L = 0.10$.
(b)~Frequency-magnitude statistics for two values of the weakening parameter: $w=w_c$ and $w>w_c$. The frequency of events of area $A$, is plotted as a function of area $A$.  At the critical point $w=w_c=0.012$ there are no system-wide events and the smaller events correlate well with equation~\ref{eqn1} taking $\tau - 1 = 1.47 \pm .05$.  Above the critical point $w = 0.04$ there are system-wide events occurring regularly producing the spike at $A = 262,144$, while the smaller events continue to obey the GR scaling relationship with $\tau$ very near $2.5$.
}
\label{damageEvents}
\end{figure}

\subsection{Threshold damage model}
Our damage model uses the cellular automaton dynamics with a failure threshold $\sigma_F$ for the initial failure and slip of all blocks and a residual stress $\sigma_R$ drawn from a random uniform distribution, as described above. We include damage into our model by considering any block that fails during an event to be weak or ``damaged'' for the remainder of the event.  The first failure of all blocks occurs at an initial failure threshold of $\sigma_F$.  Any subsequent failure of a damaged block (one which has slipped once or more during the current event) will occur at a reduced failure threshold value $\sigma_F*(1-w)$, where $0<w<1$. The amount of damage to a block is controlled by the input weakening parameter $w$. Following termination of the avalanche of slipping blocks, the initial failure stress threshold of each block is reset to the original value $\sigma_F$.  Thus we have damage, or weakening of the system following initial slip of each block by the ratio $w$, followed by healing back to the original failure threshold $\sigma^{F}$ at the termination of the avalanche (macroscopic sliding event).  In the following results, we investigate how the weakening parameter $w$ effects the behavior of the system, particularly how beyond the critical value $w_c$, the weakening parameter acts as a scaling field for the probability of system-wide events to occur.

We next illustrate the behavior of our damage model when the weakening parameter $w$ is greater than the critical value $(w>w_c)$ so that system-wide events occur regularly.  In Fig.~\ref{damageEvents}(a) the rupture areas $A$ are given as a function of event number (natural time) in Fig.~\ref{damageEvents} for $w = 0.04~(w_{c} = 0.012)$.  System-wide events occur quasi-periodically in natural time.  

In Fig.~\ref{damageEvents}(b) we give frequency-magnitude (Gutenburg-Richter) distribution for cluster sizes $A$ for two values of the weakening parameter, $w=0.012=w_c$ and $w=0.04>w_c$. The frequency of events of area $A$, is given as a function of area.  For $w=w_c$, shown with boxes, there are no system-wide events.  For $w > w_c$, shown with circles, system-wide events occur regularly, producing the large spike at area $A=262,144$.  The system-wide events have a different structure than the smaller events.  The smaller events have a diffuse structure with a mixture of failed and unfailed blocks where each failed block typically slips only once during the event.  The system-wide events are dense with all blocks failing and each failed block is likely to slip multiple times during the event.  Also included in this figure is the correlation between the $w=w_c$ data and the straight-line relation given in equation~\ref{eqn1}.  Good agreement is obtained by fitting the region from $A = 1 - 1000$ giving a scaling exponent $\tau - 1 = 1.47\pm .05$.  The system parameters are the same as in~\ref{Events}. Similar results are obtained for simulations using a larger lattice sizes $L = 768$ and $1024$.

\section{Simulation Results}
The primary purpose of our simulations is to study the role of the weakening parameter $w$. All simulations will assume a failure stress $\sigma_F = 100$, a residual stress $\sigma_R = 25 \pm 25$, and a loader spring constant $K_L = 1$ and periodic boundary conditions.  For no weakening (base model), $w = 0$, system-wide clusters of failed sites in slider-block models are exceedingly rare, and appear only in the mean-field limit as $qK_C\rightarrow \infty$, or alternatively as the range of interaction $R \rightarrow \infty$~\citep{Gras1994}.  In our simulations, we find that as the weakening parameter $w$ increases $( w > 0 )$, a value is reached at which system-wide events (``characteristic earthquakes'') begin to appear.  Indeed, this result was noted for a different model in~\citep{BenZ1999}, which demonstrated that it is possible to tune such a model so that intermittency in the appearance of system-wide avalanches is observed.  This suggests that we view the existence of system-wide events of failed sites in our slider-block model in a similar way to the infinite cluster in percolation, namely, that system-wide events should be associated with an order parameter and that there are associated critical exponents.  

Since the weakening parameter $w$ controls the nature of the scaling in this model, we conjecture that $w$ plays a similar role to the occupation probability $p$ in site percolation.  Thus we define an order parameter $P(w)$ to be the ensemble average of the probability of a site (block) selected at random being part of a system-wide cluster (``characteristic event'').  Here we consider system-wide events to be those that have a total area equal to the lattice area.  The order parameter we have defined is analogous to the `strength of the infinite network' order parameter $P(p)$ defined in percolation theory~\citep{Stau1994}.  By defining a ``characteristic events'' as encompassing the entire lattice area, we can calculate $P(w)$ simply as the fraction of observed avalanche events that are system-wide events during the simulation run.  In general we find that for $w < w_c$, $P(w) = 0$ in any finite time interval, but that for some value, $P(w) >0$.

The value of $w_{c}$ is determined numerically and is equal to the value of $w$ beyond which system-wide events (``characteristic earthquakes'') begin to appear.
We have determined the critical value of the weakening parameter $w_{c}$ when system-wide events (``characteristic earthquakes'') begin to appear is $w_{c} = 0.012$ for the simulation parameters used in Figs.~\ref{Events} and~\ref{damageEvents}. As discussed above we define the order parameter $P(w-w_{c})$ as the ensemble average of the probability a site selected at random is part of the system-wide event, where a system-wide event encompasses then entire lattice.  We calculate the order parameter as the fraction of observed avalanche events that are system-wide events.  Near a critical point we would expect to find that:
\begin{equation}
P(w-w_c) \propto (w-w_{c})^{\beta}.
\label{scaling}
\end{equation}

\begin{figure}
\centering
\includegraphics[width = 0.7\columnwidth]{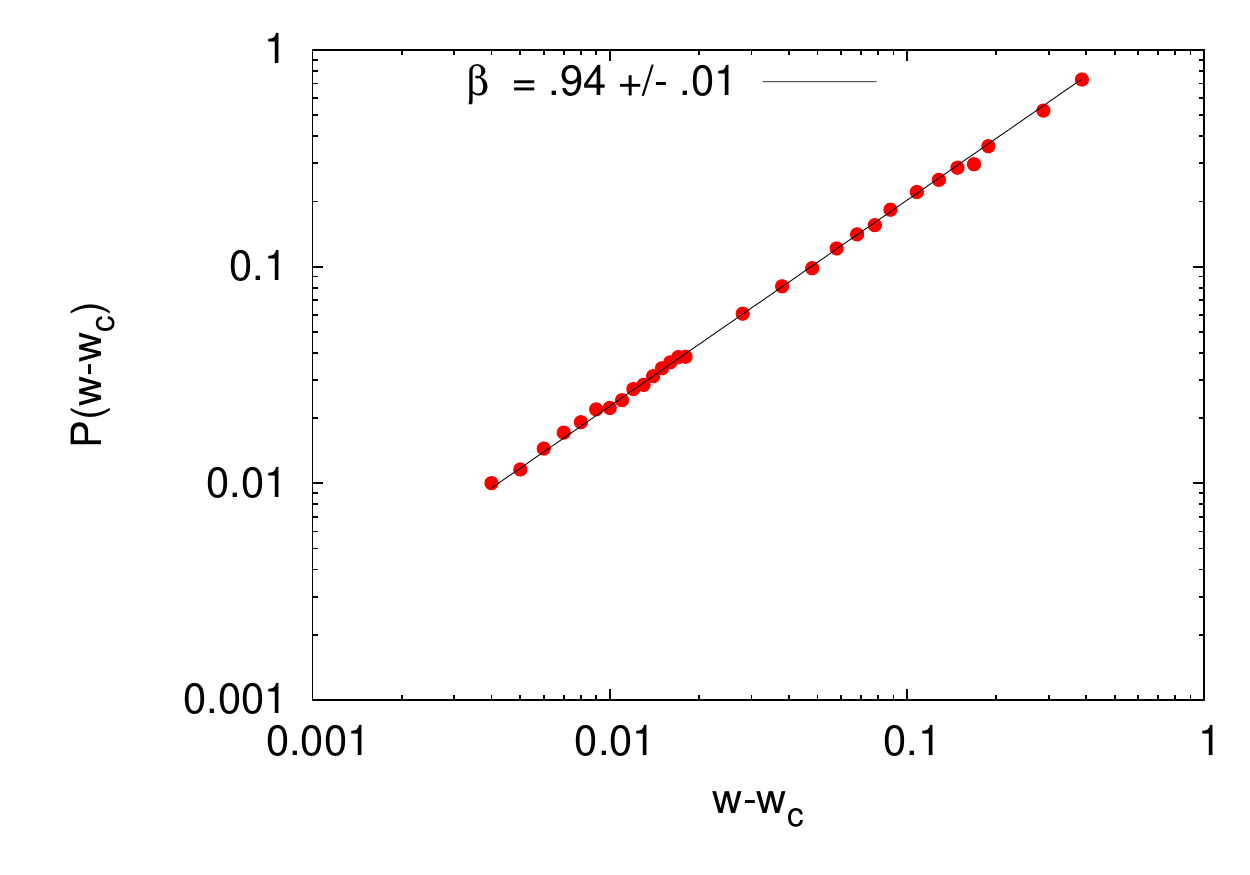}
\caption{The order parameter vs the weakening value adjusted by the critical weakening value.  The order parameter is defined as the probability of of a system wide event occurring.  The plot is on a logarithmic scale emphasizing the power-law dependence of $P(w-w_{c})$ with a scaling exponent very near $\beta = 1$. The parameters of the model are the same as in Fig.~\ref{Events}.
}
\label{betaplot}
\end{figure}

In Fig.~\ref{betaplot} we plot the dependence of $\log P$ on $\log (w-w_{c})$ for the same system parameters of the model used in Fig.~\ref{Events}.  Each data point represents a simulation run with differing value of $w$.  The range of $w$ is bounded on the right by the system that has all events spanning the entire lattice, that is $P(w-w_c) = 1$.
The value of scaling exponent for the initial scaling region is near $1$, the same value that characterizes mean-field percolation as defined in equation~\ref{scaling}. The observed $\beta = 0.94$.  As we increase the system size, smaller values of $w$ yield non-zero probabilities of system-wide events, extending the scaling region to the left.  In addition, the larger system sizes produce a slightly larger value of the scaling exponent $\beta$ with an apparent asymptotic value of $\beta = 1$.  We have measured the scaling exponent for systems with lattice dimension of $256, 512$ and $768$.

\section{Analysis}
\subsection{Scaling in non-equilibrium systems}
It is well known that high dimensional models composed of interacting sites, spins, trees, cells, units, or blocks demonstrate clustering and scaling phenomena~\citep{Gold1992, Mala2000, Bak1988, Clar1996}.    Ising models, which are often used to demonstrate both equilibrium phase transitions (second order) and non-equilibrium transitions (nucleation), have been extensively analyzed~\citep{Gold1992}.    More recently, it has been observed that non-equilibrium systems, including sandpile (self-organized critical) and forest fire models~\citep{Bak1988, Clar1996} also exhibit clustering and scaling phenomena.  For the case of slider-blocks, it was found that the energetics of the model in the near mean-field regime lead to a Boltzmann factor for the energy fluctuations~\citep{Rund1995, Klei1997, Tiam2003}.  The Boltzmann distribution of energy fluctuations, along with the apparent punctuated ergodicity as measured by the TM metric, described below, imply that the scaling properties of these models may have a similar origin to scaling properties in equilibrium systems. Later, it was found that the same ergodic property holds for coupled map lattices in the near mean-field regime~\citep{Egol2000}.

For Ising models, it has been known for many years~\citep{Coni1980} that the scaling exponents associated with the Ising critical point can be mapped onto a percolation model constructed from a correlated site-bond probability.  By themselves, random-site percolation clusters are not isomorphic to Ising clusters, because Ising clusters have both a correlated component, as well as a random (geometric) component.  However, it was shown~\citep{Coni1980} that a bond probability can be constructed to define bond percolation clusters from the Ising spin clusters that are in fact isomorphic to random site percolation clusters.  The critical exponents are the same, and the percolation critical point maps onto the Curie temperature defining the ferromagnetic-paramagnetic transition.  Thus the bond probability is used to separate the correlated and geometric effects that define the Ising clusters.  

For slider-block models, the physics is different but related, and in fact, simpler.  Here a block fails only if the stress is large enough.   So as~\citet[pg. 64]{Klei2000} note, all clusters of failed blocks are constructed from correlated components, and there is no need for an additional bond probability.  For that reason, we expect that the scaling exponents should map directly onto those for an associated site percolation problem.  Since we are dealing with a near mean-field model having long range interactions, this implies that the scaling exponents for the slider-block model should be the same as the Bethe lattice (mean-field percolation), as the range of interaction of the slider-block model becomes large.  Determining the validity of this correspondence is  a major motivation for the results presented here.

We have shown in Figs.~\ref{Events}(b) and~\ref{damageEvents}(b), the frequency-magnitude scaling for the slider-block model obeys the scaling relation given in~\ref{eqn1}.  In Fig.~\ref{betaplot} we have also indicated the scaling relation between the order parameter $P$ and the scaling field $w-w_c$ introduced into our damage model.  The values $\tau - 1 = 1.51$ and $\beta = .94$, provided we are using a finite lattice, are within experimental error of the corresponding mean-field percolation model scaling exponents.

\subsection{Ergodic Dynamics}

The slider-block model is an example of a driven nonlinear threshold system where the dynamics of the system are strongly correlated in space and time. It has been shown that as the range of interaction becomes large, the slider-block model approaches a mean-field spinodal resulting in scaling phenomena, such as the frequency-magnitude (Gutenburg-Richter) shown in Fig~\ref{Events}(a)~\citep{Klei2000}.  In this limit, the system resides in a metastable equilibrium, where equilibrium-like properties, such as stationary dynamics, equally probable microstates, and effective ergodicity all hold true.  This metastable state is disrupted by large events, and the system evolves to a new metastable state.  Therefore, the slider-block model displays punctuated or intermittent ergodicity~\cite{Ferg1999}.  

\begin{figure}
\centering
\mbox{
\subfigure{\includegraphics[width = 0.45\columnwidth]{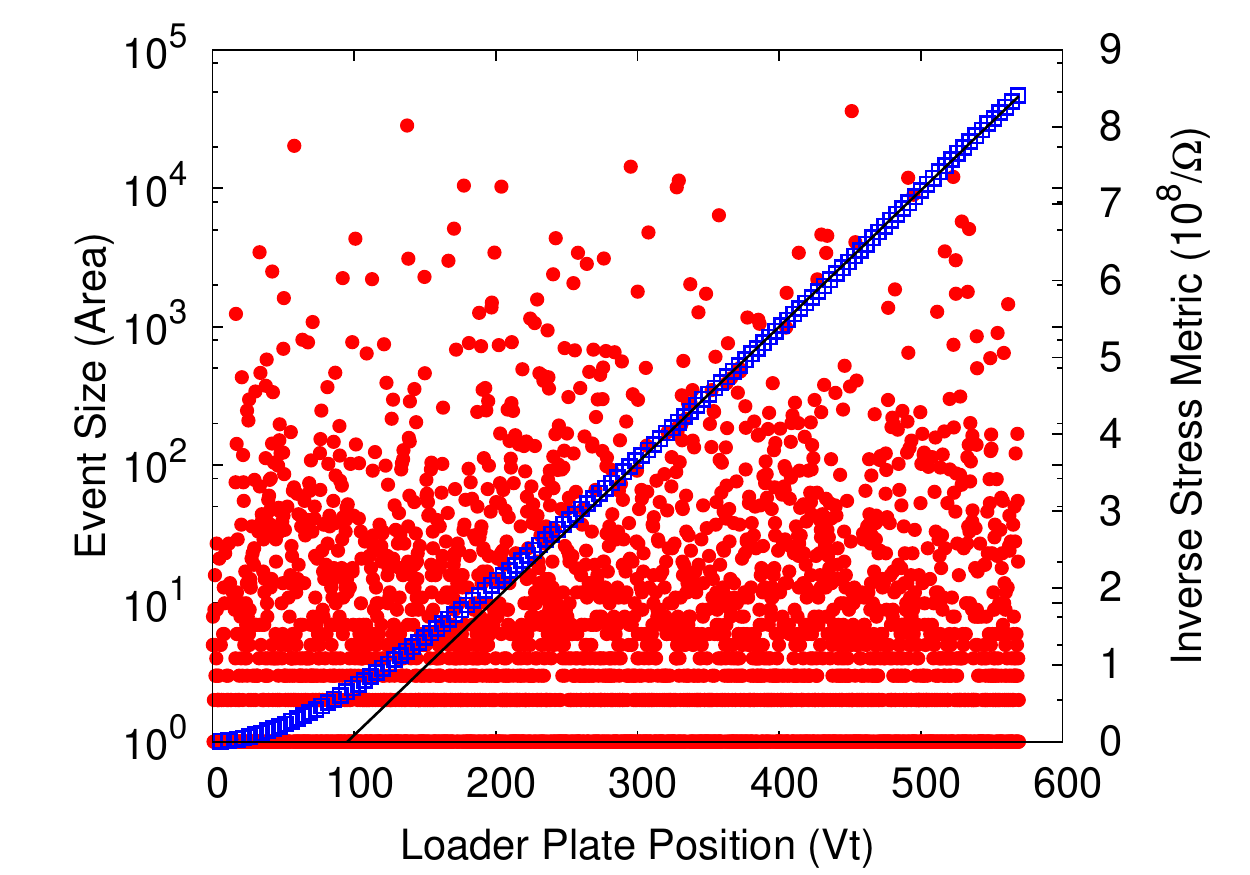}}
\quad
\subfigure{\includegraphics[width = 0.45\columnwidth]{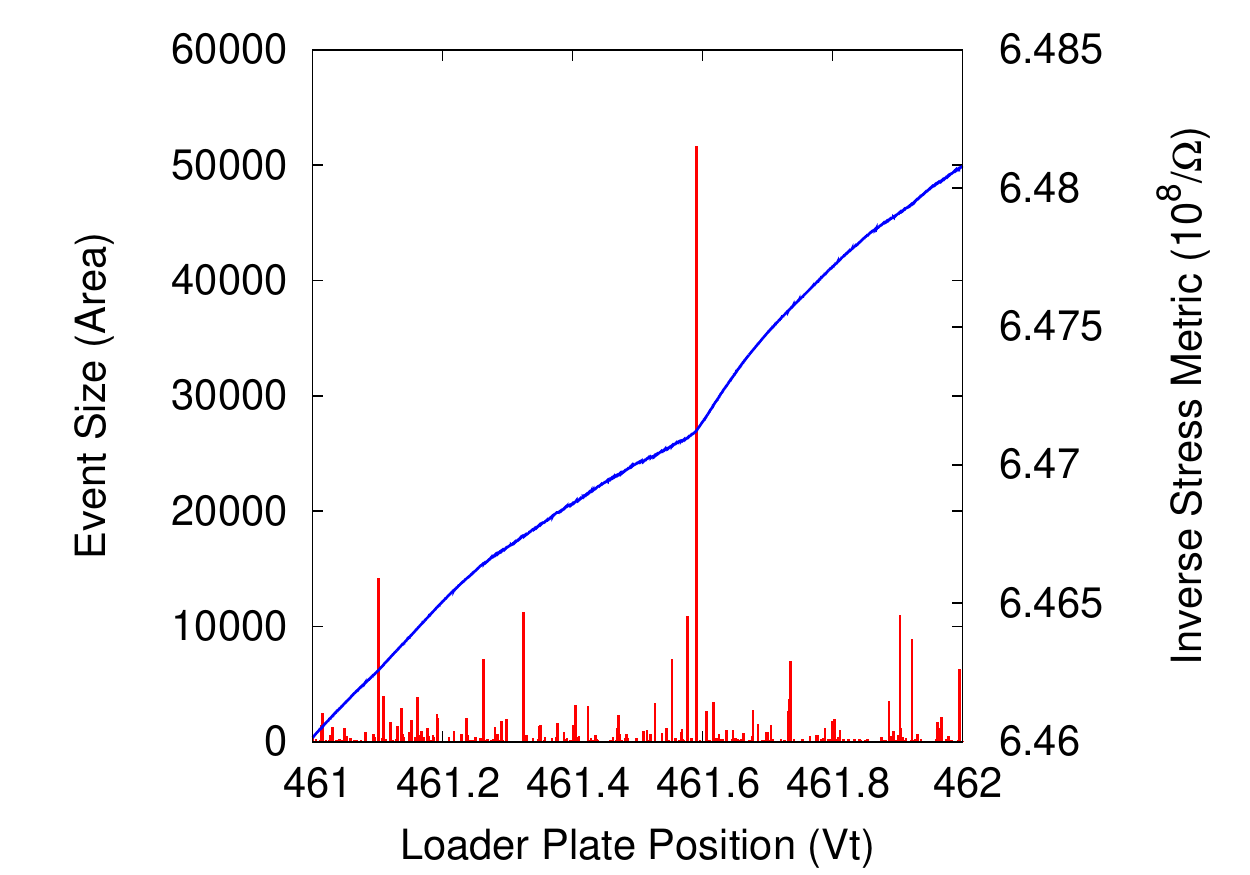}}}
\caption{The inverse TM metric for measuring effective ergodicity is overlayed with the event size sequence as a function of loader plate position, for the "threshold damage model" at the critical point $w=w_c$.  A system is effectively ergodic if the stress metric decreases as $1/t$. In (a) we plot the inverse TM metric vs loader plate position (time). This plot shows a linear trend, displaying the system is effectively ergodic from $Vt =300 -600$. In (b), we show a detailed view of a break in the ergodicity caused by a large event.}
\label{Ergodicity}
\end{figure}

\begin{figure}
\centering
\includegraphics{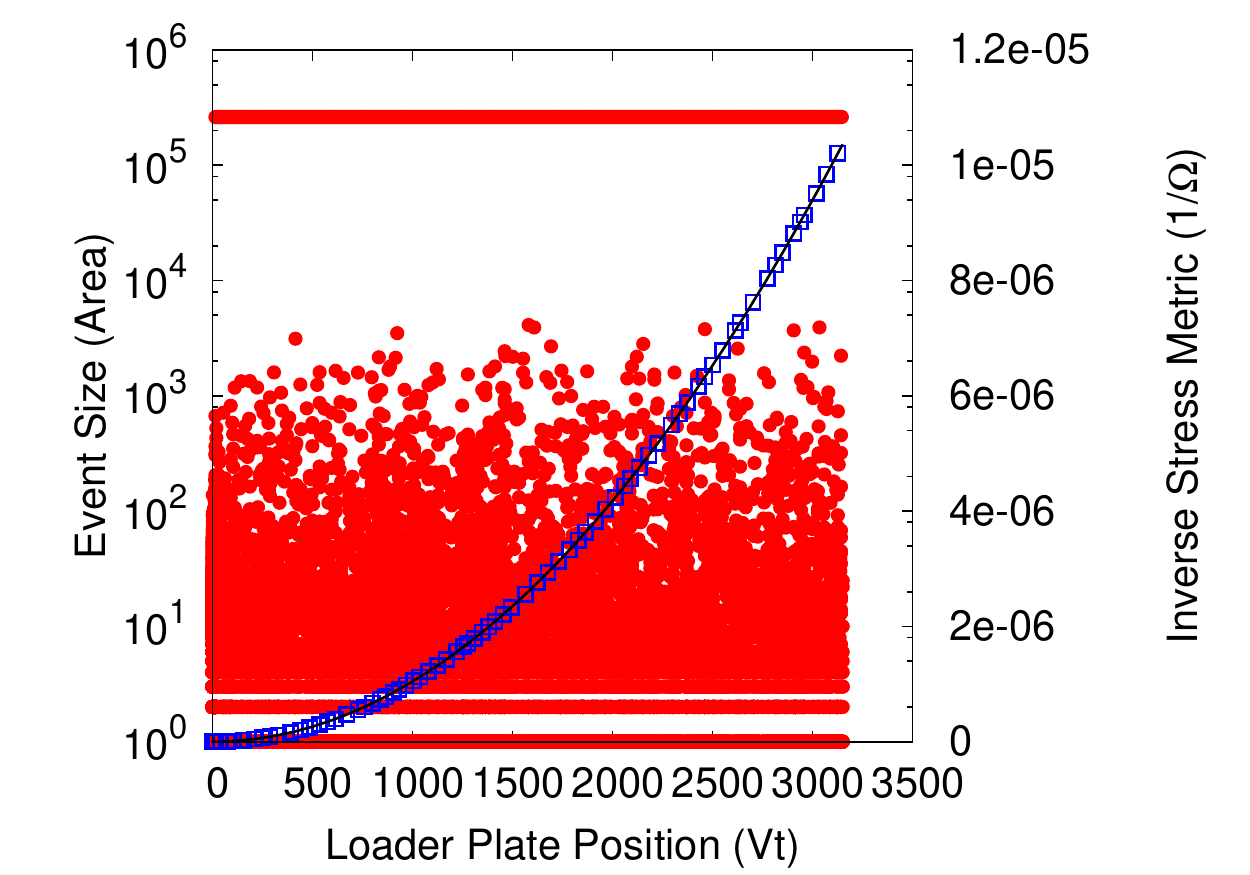}
\caption{The inverse TM metric and the event size sequence are shown for the "threshold damage model" above the critical point $(w>w_c)$.  The inverse metric shows a $1/t^2$ trend, displaying that this model is not ergodic over the observed time period.  The grid-size events, shown with "dots", prevent the system from residing in a meta-stable equilibrium for any extended period of time.}
\label{Ergodicity2}
\end{figure}

A measure of ergodicity can be obtained using the Thirumalai-Mountain (TM) fluctuation metric~\citep{Thir1990, Thir1993}.  The TM metric, $\Omega(t)$, measures the difference between the time average of an observable at each site and its ensemble average over the entire system.  The TM metric relies on the idea of statistical symmetry, where the statistics of one particle (block) is presumed identical to those of the entire system.  Here the observable is the stress on each block $\sigma(t)$.  The metric is defined as follows:

\begin{equation}
\Omega(t) = \frac{1}{N}\sum_i[\bar{\sigma}_i(t) - \langle\bar{\sigma}(t)\rangle]^2
\label{metric1}
\end{equation}
where $N$ is the number of blocks, and the quantities $\bar{\sigma}_i(t)$ and $\langle\bar{\sigma}(t)\rangle$ are given by

\begin{equation}
\bar{\sigma}_i(t) = \frac{1}{t}\int_0^t dt'{\sigma}_i(t')
\label{metric2}
\end{equation}
and
\begin{equation}
\langle\bar{\sigma}(t)\rangle = \frac{1}{N}\sum_i \bar{\sigma}_i(t).
\label{metric3}
\end{equation}
If the system is effectively ergodic at long times, the TM metric decreases in time as, $\Omega(t) = \frac{D_e}{t}$~\citep{Thir1990}, where $D_e$ is a diffusion constant related to the rate at which the phase space is explored.  In Fig.~\ref{Ergodicity}, we plot the inverse of stress metric introduced in~\citep{Klei2000}.  As noted above, a system is effectively ergodic if the stress metric decreases as $1/t$, therefore, a plot of the inverse metric, $\frac{1}{\Omega}$, vs time, $t$, will show a linear relationship for ergodic systems. In Fig.~\ref{Ergodicity}(a), we plot the inverse stress metric, with ``boxes'', and the sequence of events, with ``dots'', as a function of the loader plate position (time) for the ``threshold damage model'' at the critical point ($w = w_c$).   The inverse TM metric shows a linear trend from loader plate position $300 - 600$.  The initial curved portion of the plot is due to transient effects. In Fig.~\ref{Ergodicity}(b), we show a small section of the plot in Fig.~\ref{Ergodicity}(a) centered on a large event of size $A\approx 50000$.  Here we show that large events disrupt the ergodicity of the model for a short period of time as evident by the ``kink'' in the inverse TM metric.  This model has punctuated ergodicity.  In Fig.~\ref{Ergodicity2} we plot the inverse metric for the ``threshold damage model'' above the critical point.  Here the system is not ergodic over any time period.  The grid size events prevent the system from residing in a metastable equilibrium for a long period of time.  The inverse metric is well fit to a quadratic function as noted in~\citep{Thir1990} for systems that are non-ergodic.  This result is in agreement with a different adaptation of ``damage'' into the long-range slider-block model proposed by~\citep{Seri2009}.  Here blocks in the lattice are removed and no longer contribute to the dynamics of the system after a specified number of failures.  At the onset of the blocks becoming ``damaged'' the system loses its ergodicity.

\section{Discussion}

\subsection{Application to Data}
It is useful to compare the scaling exponents determined from our numerical simulations with scaling exponents seen in laboratory experiments and in nature.  In all of these experiments, the fracturing and acoustic emissions are mixed-mode, not simple shear sliding as is the case for our simulations.  As a result, we expect to observe a range of values for scaling exponents.  Typically, the data are in the form of probability density function (PDF) as a function of energy E of acoustic emissions (AE).  For frequency vs. energy measurements, and assuming that area of microcrack is proportional to energy released, which would be the case if crack area is proportional to crack energy for these microcracks, a range of scaling exponents $\tau - 1$ are observed, from 1.2 to 2.0, with most values clustering around 1.5 as is observed in our simulations~\citep{Mein2008,Maes1998,Houl1996,Garc1997}.  On the other hand, if crack energy released is proportional to crack area to the 3/2 power, as is the case for macroscopic cracks, then we would expect scaling exponents at the lower end of the observed range, clustering around $\tau - 1 = 1.0$.

\subsection{Summary}
In this paper, we have discussed the various types of simulations used to model the physics of material deformation and damage, including fiber-bundle models and slider-block models.  We have also discussed the experimental and observational motivation for incorporating damage and healing into these models.  We infer that healing must occur on faults where stick-slip events (earthquakes) occur regularly, motivating the addition of damage and partial healing into a traditional earthquake model.  We included the concept of damage and healing into the standard cellular-automata slider-block model by means of failure threshold weakening.  We reduced (``damaged'') each block that slips during an event by allowing subsequent failures of the damaged blocks to occur at a lower failure threshold parameterized by $w$.  After the event has terminated we heal all blocks to their original failure threshold levels.

We have shown that the weakening parameter $w$ behaves as a scaling field for the ensemble average probability of a site selected at random to be part of a system-wide event $P(w-w_c)$, just as the occupation probability is a scaling field for the strength of the infinite network in percolation theory. We also discussed the mean-field percolation problem along with the various critical exponents that can be measured. The results of our simulation yield scaling exponents that approach those of the in Bethe lattice (mean-field) percolation problem, specifically $\beta = 1$.  We have demonstrated that the ``threshold weakening model'' displays ergodic behavior up to the critical point.  Beyond the critical point $w_c$, the model is no longer ergodic.  We conclude that ``damage'' as we have modeled it, is a non-equilibrium process. We have also shown the scaling exponents found in our slider-block model to be within the range of those found in experiments.  We conclude that the mean-field interactions may dominate the failure of material. Further studies of this model including an investigation of time dependent healing are currently being studied.
\section*{Acknowledgements}
Research by  of J.D.G., J.B.R., J.R.H., and D.L.T. has been supported by a grant from the U.S. Department of Energy, Office of Basic Energy Sciences to the University of California, Davis, DE-FG03-95ER14499.  The research by W. Klein has been supported by a grant from the U.S. Department of Energy, Office of Basic Energy Sciences to Boston University, DE-FG02-95ER14498.

\section*{References}
\bibliography{/home/joe/Documents/references/BIB}
\bibliographystyle{model1-num-names}

\end{document}